\documentclass[
aps,prd,
10pt,
superscriptaddress,
amsfonts,amssymb,amsmath,
preprintnumbers,
nofootinbib,
showpacs,
a4paper,
eqsecnum,
]{revtex4}

\usepackage{graphicx}

\begin{document}

\preprint{YITP-10-105}

\title{
Inflation with a Weyl term, or ghosts at work
}

\author{Nathalie Deruelle}
\affiliation{
APC, UMR 7164 du CNRS, Universit\'e Paris 7,
75205 Paris Cedex 13, France
}
\author{Misao Sasaki}
\affiliation{
Yukawa Institute for Theoretical Physics, Kyoto University,
Kyoto 606-8502, Japan
}
\affiliation{
Korea Institute for Advanced Study,
Seoul 130-722, Republic of Korea
}
\author{Yuuiti Sendouda}
\affiliation{
Graduate School of Science and Technology, Hirosaki University,
Hirosaki 036-8561, Japan
}
\affiliation{
APC, UMR 7164 du CNRS, Universit\'e Paris 7,
75205 Paris Cedex 13, France
}
\author{Ahmed Youssef}
\affiliation{
APC, UMR 7164 du CNRS, Universit\'e Paris 7,
75205 Paris Cedex 13, France
}

\begin{abstract}
In order to assess the role of ghosts in cosmology, we study the evolution of linear cosmological perturbations during inflation when a Weyl term is added to the action.
Our main result is that vector perturbations can no longer be ignored and that scalar modes diverge in the newtonian gauge but remain bounded in the comoving slicing.
\end{abstract}


\maketitle

\section{
Introduction
}

Consider the action
\begin{equation}
S
= \frac{1}{2 \kappa} \int\!d^4x\,\sqrt{-g}\,R
  - \frac{1}{2}
    \int\!d^4x\,\sqrt{-g}\,
    \left(\partial_\mu \phi\,\partial^\mu \phi + 2 V(\phi)\right)
  - \frac{\gamma}{4 \kappa}
    \int\!d^4x\,\sqrt{-g}\,C_{\mu\nu\rho\sigma}\,C^{\mu\nu\rho\sigma}\,,
\label{eq:action}
\end{equation}
where $ g $ is the determinant of the metric $ g_{\mu\nu} $, $ R $ is the scalar curvature and $ C_{\mu\nu\rho\sigma} $ is the Weyl tensor.\footnote{\label{fn:def}
Units: $ \kappa = 8 \pi\,G $, $ c = 1 $; $ \gamma $ has dimension $ \text{length}^2 $.
$ \kappa $ has dimension $ \text{length}/\text{mass} $.
Conventions: $ (-+++) $;
$ R^\mu{}_{\nu\rho\sigma} = \partial_\rho \Gamma^\mu_{\nu\sigma} - \partial_\sigma \Gamma^\mu_{\nu\rho} + \cdots $;
$ R_{\nu\sigma} = R^\mu{}_{\nu\mu\sigma} $;
$ R = g^{\mu\nu} R_{\mu\nu} $;
$ G_{\mu\nu} = R_{\mu\nu} - \frac{1}{2} g_{\mu\nu} R $.
$ C_{\mu\nu\rho\sigma} = R_{\mu\nu\rho\sigma} - \frac{1}{2} (g_{\mu\rho} G_{\nu\sigma} - g_{\mu\sigma} G_{\nu\rho} - g_{\nu\rho} G_{\mu\sigma} + g_{\nu\sigma} G_{\mu\rho}) - \frac{R}{3} (g_{\mu\rho} g_{\nu\sigma} - g_{\mu\sigma} g_{\nu\rho}) $.
Greek indices run from $ 0 $ to $ 3 $; latin indices run from $ 1 $ to $ 3 $.
}

The two first terms describe Einstein's gravity minimally coupled to a scalar field $ \phi $ with potential $ V(\phi) $ (and can also be seen as the Einstein frame formulation of $ f(R) $ theories of gravity, see \cite{Bicknell}).
The last term was first introduced by Weyl \cite{Weyl} (see \cite{Schmidt} for a review of the early literature) and has ever since been present on the market of gravity theories, either, in recent decades, as a quantum correction popping up from various theories of quantum gravity (starting with \cite{Utiyama}) or, more recently, as a phenomenological modification of Einstein's General Relativity to account for e.g.\ dark matter or energy, see e.g.\ \cite{Mannheim}.

Extremisation of \eqref{eq:action} with respect to the metric yields the equations of motion:
\begin{equation}
G_{\mu\nu} - \gamma\,B_{\mu\nu}
= \kappa\,T_{\mu\nu}
\quad\text{with}\quad
\left\{
\begin{aligned}
T_{\mu\nu}
&
= \partial_\mu \phi\,\partial_\nu \phi
  - g_{\mu\nu}\,
    \left(\frac{1}{2} \partial_\rho\phi\,\partial^\rho\phi + V(\phi)\right)\,, \\
B_{\mu\nu}
&
= 2 D^\rho D^\sigma C_{\mu\rho\nu\sigma}
  + G^{\rho\sigma}\,C_{\mu\rho\nu\sigma}\,,
\end{aligned}
\right.
\label{eq:eoms}
\end{equation}
where $ G_{\mu\nu} $ is the Einstein tensor and $ B_{\mu\nu} $ is the Bach tensor \cite{Bach}.
The divergence of the left-hand-side being identically zero (generalized Bianchi identity), the Klein-Gordon equation for the scalar field is redundant.
The structure of this action and equations of motion has been thoroughly studied, in particular their hamiltonian formulation, see \cite{Boulware}, as well as some of their solutions, for example those of Bianchi type I, see \cite{Schmidt2}.

Equations \eqref{eq:eoms} are fourth order differential equations for the metric components.
They therefore possess extra, ``run-away'', solutions compared to the Einstein, $ \gamma = 0 $, ones, which can drastically modify the predictions, even in the small $ \gamma $ limit.

Indeed, as shown in \cite{Stelle}, the theory possesses ghosts when linearised around Minkowski spacetime, that is, the hamiltonian contains negative kinetic terms and, as a consequence, the energy spectrum of the metric perturbations is not bounded from below (just as in the toy model with lagrangian $ (\square \phi)^2 $ studied earlier by Pais and Uhlenbeck, \cite{Pais}).
In fact most ``higher derivative theories'', that is, yielding equations of motion of differential order higher than two, are thought to possess ghosts (to the notable exception of $ f(R) $ theories of gravity, see \cite{Bicknell}).

Although the presence of these ghost degrees of freedom is harmless at linear level around Minkowski spacetime on which all modes propagate independently of each other (see e.g.\ \cite{Bogdanos}), there are strong arguments to predict that they yield a catastrophic collapse of any system when coupled to other fields, their energy running down to minus infinity in a finite time (see e.g.\ \cite{Pais,Cline}).
However, since the introduction of coupling implies that the equations of motion become non-linear, this catastrophic behaviour has been explicitly exhibited on toy models only, see e.g.\ \cite{Smilga}.
By the same token, most proposals to tame ghosts have been also illustrated by toy models only, see e.g.\ \cite{Hawking,Bender}.
Showing explicitly how the ghosts present in the particular theory of gravity described by (\ref{eq:action}--\ref{eq:eoms}) may render it unviable when self-coupling or coupling to external fields is introduced, has not been done so far.

Now it may happen that the malignancy of ghosts shows up already at linear level, if the background is richer than Minkowski spacetime.
However little has been done in this direction.
In \cite{Clunan} the Hawking-Hertog proposal \cite{Hawking} was used to tame the tensor ghosts on a de Sitter background.
In \cite{Nelson} the equations of motion for the scalar perturbations on a Friedmann-Lema\^itre background were spelt out but not thoroughly analysed.

In this paper, we aim at assessing the role of the Weyl term on the evolution of linear cosmological perturbations when the Friedmann-Lema\^itre background is that of single-field inflation.
After having obtained the equations of motion for the perturbations, as well as the action from which they derive, we analyse the evolution of the modes.
We see that tensor modes are not drastically modified by the presence of ghosts.
Vector modes (which have been so far ignored in the literature) are no longer absent as in Einstein's theory but do propagate when the Weyl term is present.
Finally we give a master equation for the evolution of all scalar modes and find that their evolution is highly gauge dependent:
they are unstable in the newtonian gauge but decay in the comoving slicing.
We conclude on what should be done next.

\section{
Cosmological background
}

When the metric is that of a conformally flat Friedmann-Lema\^itre spacetime, $ ds^2 = a(\eta)^2\,(-d\eta^2 + d\vec x^2) $, the Weyl term does not contribute and the equations of motion \eqref{eq:eoms} for the scale factor $ a(\eta) $ and the background inflaton $ \phi = \varphi(\eta) $ are:
\begin{equation}
\frac{\kappa}{2} \varphi'^2
= \mathcal H^2 - \mathcal H'\,,
\quad
\kappa\,a^2\,V
= 2 \mathcal H^2 + \mathcal H'\,,
\end{equation}
where a prime denotes differentiation with respect to conformal time $ \eta $ and where $ \mathcal H \equiv a'/a $.
The Klein-Gordon equation which entails,
\begin{equation}
\varphi'' + 2 \mathcal H\,\varphi' + a^2\,V_{,\varphi}
= 0\,,
\end{equation}
where $ V_{,\varphi} \equiv \frac{dV}{d\phi}|_{\phi=\varphi} $, is also useful.

When $ V(\phi) \propto \phi^{2 n} $ these equations are those of chaotic inflation \cite{Linde} and a detailed analysis of their solutions can be found in e.g.\ \cite{Mukhanov}.
In a nutshell: after a transitory period the scale factor increases quasi-exponentially in cosmic time $ t = \int^\eta a\,d\eta $, while the scalar field slowly decreases.
At the end of inflation the scalar field oscillates and settles at the bottom of its potential well while the scale factor increases in average as some power of $ t $ ($ t^{2/3} $ if $ V(\phi) \propto \phi^2 $).

\section{
Equations of motion
}

Consider the perturbed metric
\begin{equation}
ds^2
= a(\eta)^2\,
  [-(1 + 2 A)\,d\eta^2
   + 2 B_i\,dx^i\,d\eta
   + (\delta_{ij} + h_{ij})\,dx^i\,dx^j]
\end{equation}
and perform the scalar-vector-tensor decomposition \cite{Bardeen}
\begin{equation}
B_i
= \partial_i B
  + \bar B_i\,,
\quad
h_{ij}
= 2 C\,\delta_{ij}
  + 2 \partial_{ij}\,E
  + \partial_i \bar E_j
  + \partial_j \bar E_i
  + \bar h_{ij}
\end{equation}
with $ \partial_i \bar B^i = \partial_i \bar E^i = \partial_i \bar h^{ij} = \bar h^i{}_i = 0 $.
In an infinitesimal coordinate transformation the following six quantities
\begin{equation}
\Psi_\mathrm n
= A + \mathcal H\,(B-E') + (B-E')'\,,
\quad
\Phi_\mathrm n
= C + \mathcal H\,(B-E')\,,
\quad
\bar\Psi_i
= \bar B_i - \bar E_i'\,,
\quad
\bar h_{ij}
\label{eq:gi}
\end{equation}
are invariant \cite{Bardeen}.
As for the perturbation $ \delta\phi $ of the scalar field it is such that
\begin{equation}
\chi_\mathrm n
= \delta\phi + \varphi'\,(B-E')
\label{eq:gi2}
\end{equation}
is gauge invariant.

We choose to work in the coordinate system $ B = E = 0, \bar E_i = 0 $ (``newtonian'' or ``longitudinal'' gauge \cite{Mukhanov}, hence the subscript $ \mathrm n $ appended to the gauge invariant scalar perturbations), so that the perturbed metric and scalar field reduce to
\begin{equation}
ds^2
= a(\eta)^2\,
  \{
   -(1 + 2 \Psi_\mathrm n)\,d\eta^2
   + 2 \bar\Psi_i\,d\eta\,dx^i
   + [(1 + 2 \Phi_\mathrm n)\,\delta_{ij} + \bar h_{ij}]\,dx^i\,dx^j
  \}\,,
\quad
\delta\phi
= \chi_\mathrm n.
\end{equation}

The necessary ingredients to expand the equations of motion \eqref{eq:eoms} at linear order in the perturbations are given in Appendix~\ref{app:pert}.
The result is:
\begin{align}
\square \bar h_{ij} - 2 \mathcal H\,\bar h_{ij}'
&
= \frac{\gamma}{a^2} \square \square \bar h_{ij}\,,
\label{eq:eoms-tensor} \\
\bar\Psi_i
&
= \frac{\gamma}{a^2} \square \bar\Psi_i\,,
\label{eq:eoms-vector}
\end{align}
and
\begin{equation}
\left\{
\begin{aligned}
6 \mathcal H\,\Phi_\mathrm n - 2 \triangle \Phi_\mathrm n
- \kappa
  \left(
   \varphi'\,\chi_\mathrm n'
   + 2 a^2\,V\,\Psi_\mathrm n
   + a^2\,V_{,\varphi}\,\chi_\mathrm n
  \right)
&
= \frac{2 \gamma}{3 a^2} \triangle \triangle W\,, \\
\mathcal H\,\Psi_\mathrm n
- \Phi_\mathrm n'
- \frac{\kappa\,\varphi'\,\chi_\mathrm n}{2}
&
= \frac{\gamma}{3 a^2} \triangle W'\,, \\
-(\Phi_\mathrm n + \Psi_\mathrm n)
&
= \frac{\gamma}{a^2} \left(W'' - \frac{1}{3} \triangle W\right)\,,
\end{aligned}
\right.
\label{eq:eoms-scalar}
\end{equation}
where $ W \equiv \Psi_\mathrm n - \Phi_\mathrm n $, $ \square \equiv \eta^{\mu\nu}\,\partial_{\mu\nu} $ and $ \triangle \equiv \delta^{ij}\,\partial_{ij} $.

These are seven equations for seven unknown quantities, two, Eq.~\eqref{eq:eoms-tensor}, for the two components of the tensor perturbations $ \bar h_{ij} $, two, Eq.~\eqref{eq:eoms-vector}, for the two components of the vector perturbations $ \bar\Psi_i $ and three, Eq.~\eqref{eq:eoms-scalar}, for the two scalar perturbations of the metric, $ \Phi_\mathrm n $ and $ \Psi_\mathrm n $, and the perturbation $ \chi_\mathrm n $ of the scalar field.
The first Eq.~\eqref{eq:eoms-scalar} is the $ (00) $-component of the scalar part of the field equations, the second is their $ (0i) $-components, the third is the $ (ij) $-components ($ i \neq j $) of their spatial part.\footnote{
The field equations \eqref{eq:eoms} have ten components.
The remaining three (which are the part of the spatial equations proportional to $ \delta_{ij} $) can be explicitly shown to be redundant.
}
Equation \eqref{eq:eoms-tensor} for the tensor perturbations are given in \cite{Clunan}.
Equation \eqref{eq:eoms-vector} for the vector perturbations seem to have been ignored so far.
Equation \eqref{eq:eoms-scalar} for the scalar perturbations can be found in \cite{Nelson}.

\section{
The action
}

The expansion of the Einstein part ($ \gamma = 0 $) of the action \eqref{eq:action} at quadratic order in the perturbations on a Friedmann-Lema\^itre background was first obtained in newtonian gauge in \cite{Mukhanov2}.
The expansion of the Weyl part is easy (see Appendix~\ref{app:pert}).
The result is (all spatial indices being raised with $ \delta^{ij} $):
\begin{equation}
\left\{
\begin{aligned}
\kappa\,S^{(\mathrm T)}
&
= \kappa\,S^{(\mathrm T)}_\mathrm E
  - \frac{\gamma}{8}
    \int\!d^4x\,
    (\bar h''_{ij}\,\bar h''^{ij}
     - 2 \partial_k \bar h'_{ij}\,\partial^k \bar h'^{ij}
     + \triangle \bar h_{ij}\,\triangle \bar h^{ij})\,, \\
\kappa\,S^{(\mathrm V)}
&
= \kappa\,S^{(\mathrm V)}_\mathrm E
  - \frac{\gamma}{4}
    \int\!d^4x\,
    (\partial_i \dot{\bar\Psi}_j\,\partial^i \dot{\bar\Psi}^j
     - \triangle \bar\Psi_i\,\triangle \bar\Psi^i)\,, \\
\kappa\,S^{(\mathrm S)}
&
= \kappa\,S^{(\mathrm S)}_\mathrm E
  - \frac{\gamma}{3}
    \int\!d^4x\,
    [\triangle (\Psi_\mathrm n-\Phi_\mathrm n)]^2
\end{aligned}
\right.
\label{eq:action-pert}
\end{equation}
with, see e.g.\ \cite{Mukhanov,Makino}:
\begin{equation}
\left\{
\begin{aligned}
\kappa\,S^{(\mathrm T)}_\mathrm E
&
= \frac{1}{8} \int\!d^4x\,a^2\,
  (\bar h'_{ij}\,\bar h'^{ij}
   - \partial_k \bar h_{ij}\,\partial^k \bar h^{ij})\,,
\quad
\kappa\,S^{(\mathrm V)}_\mathrm E
= \frac{1}{4} \int\!d^4x\,a^2\,
  \partial_i \bar\Psi_j\,\partial^i \bar\Psi^j\,, \\
\kappa\,S^{(\mathrm S)}_\mathrm E
&
= \frac{1}{2} \int\!d^4x\,a^2\,
  [-6 \Phi_\mathrm n'^2
   + 12 \mathcal H\,\Psi_\mathrm n\,\Phi_\mathrm n'
   + 2 \partial_i \Phi_\mathrm n\,
     (2 \partial^i \Psi_\mathrm n
      + \partial^i \Phi_\mathrm n)
   - 2 (\mathcal H' + 2 \mathcal H^2)\,\Psi_\mathrm n^2 \\
&
\quad\quad
  + \kappa\,
    (\chi_\mathrm n'^2
     - \partial_i \chi_\mathrm n\,\partial^i \chi_\mathrm n
     - a^2\,V_{,\varphi\varphi}\,\chi_\mathrm n^2
     - 6 \varphi'\,\Phi_\mathrm n'\,\chi_\mathrm n
     - 2 \varphi'\,\chi_\mathrm n'\,\Psi_\mathrm n
     - 2 a^2\,V_{,\varphi}\,\Psi_\mathrm n\,\chi_\mathrm n)]\,.
\end{aligned}
\right.
\label{eq:action-pert-Einstein}
\end{equation}

Extremisation of $ S^{(\mathrm T)} $ with respect to the tensor perturbations $ \bar h_{ij} $ readily yields the equations of motion \eqref{eq:eoms-tensor}.
Similarly the extremisation of $ S^{(\mathrm V)} $ with respect to the vector
 perturbations $ \bar\Psi_i $ yields the equations of motion \eqref{eq:eoms-vector}.

As far as we are aware the fact that the extremisation of $ S^{(\mathrm S)} $ with respect to the scalar perturbations $ \Phi_\mathrm n $, $ \chi_\mathrm n $ and $ \Psi_\mathrm n $ also yields back the equations of motion \eqref{eq:eoms-scalar} does not appear in the literature, even in the case of standard inflation when $ \gamma = 0 $.
This is however the case, as we show it in some detail in Appendix~\ref{app:fromaction}.

This shows that one can completely fix the coordinate system from start (instead of keeping the ten metric perturbations plus $ \delta\phi $), obtain the action at quadratic order in terms of seven perturbations only, and still recover the seven equations of motion after extremisation, at least when working in the newtonian gauge.
Doing so, we do not loose any algebraic constraints (or lower derivative equations if $ \gamma \neq 0 $).
There is therefore no need at linear level to keep the coordinate system unspecified (that is, keep the lapse and shift as free Lagrange multipliers) as is necessary in any hamiltonian formulation of the full theory, and as is usually done in the theory of linear cosmological perturbations, see e.g.\ \cite{Mukhanov} or \cite{Makino}.

\section{
Evolution
}

We shall work in Fourier space, that is, we expand the seven perturbations $ \bar h_{ij}(\eta,x^k) $, $ \bar\Psi_i(\eta,x^k) $, $ \Phi_\mathrm n(\eta,x^k) $, $ \Psi_\mathrm n(\eta,x^k) $ and $ \chi_\mathrm n(\eta,x^k) $, collectively denoted by $ f(\eta,x^k) $, as
\begin{equation*}
f(\eta,x^k)
= \int\!d^3k\,f_{\vec k}(\eta)\,e^{i\,\vec k \cdot \vec x}
\quad\text{with}\quad
f^*_{-\vec k}(\eta)
= f_{\vec k}(\eta)
\end{equation*}
and, to simplify notations, we shall omit the index $ \vec k $ on the Fourier component $ f_{\vec k}(\eta) $.
We recall how the Fourier components evolve in standard inflation when $ \gamma = 0 $ in Appendix~\ref{app:standard}.

When the Weyl term is present, that is when $ \gamma \neq 0 $, the structure of the equations of motion \eqref{eq:eoms-tensor} \eqref{eq:eoms-vector} \eqref{eq:eoms-scalar} changes drastically.

\subsection{
Vector perturbations
}

Let us start with Eq.~\eqref{eq:eoms-vector} for the two vector perturbations $ \bar\Psi_i $.
It is no longer an algebraic constraint as in standard inflation when $ \gamma = 0 $.
During inflation when spacetime can be harmlessly approximated by a de Sitter space with $ a = 1/(-H\,\eta) $, Eq.~\eqref{eq:eoms-vector} reads, in Fourier space
\begin{equation}
\frac{d^2}{dz^2} \bar\Psi_i
+ \left(1 + \frac{1}{\gamma\,H^2\,z^2}\right)\,\bar\Psi_i
= 0
\label{eq:eoms-vector-dS}
\end{equation}
with $ z = -k\,\eta $.
As for $ \bar\Psi_i $, it stands now for the Fourier component $ \bar\Psi_{i,\vec k}(\eta) $.
Since \eqref{eq:eoms-vector-dS} is a second order differential equation,
 there are (for each $ \vec k $) two vector degrees of freedom,
 one for each polarisation $ i $.
The two independent solutions of \eqref{eq:eoms-vector-dS} for
 each degree of freedom, that is its two modes, can be given in terms of
 Bessel (or Hankel) functions of index $ \nu = \sqrt{1/4 - 1/(\gamma\,H^2)} $,
 $ \bar\Psi_i \propto z^{1/2}\,H_\nu^{(\sigma)}(z) $ ($ \sigma = 1,2 $).

Now, $ \gamma $ must be positive otherwise $ \bar\Psi_i $ would
 behave as a tachyon on flat spacetime, see Eq.~\eqref{eq:eoms-vector}.
We also have that
\begin{equation*}
\ell_\mathrm{Planck}^2 \lesssim \gamma \ll \frac{1}{H^2}
\end{equation*}
if inflation occurs at GUT scale and if the Weyl-correction is
 due to a \emph{low}-energy approximation of some quantum gravity theory.

\begin{figure}
\includegraphics[scale=1.0]{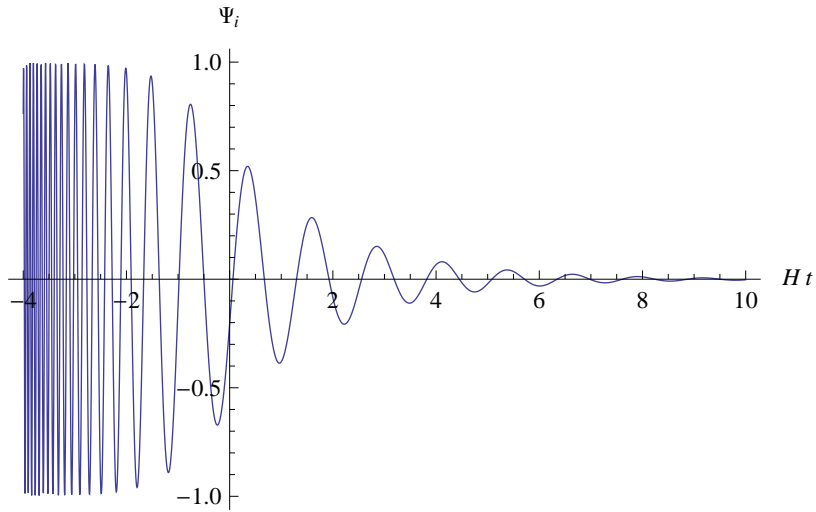}
\caption{\label{fig:vector}
Evolution in cosmic time $ t = \int^\eta\!a\,d\eta $ of a Weyl vector mode
 $ \Re[\bar\Psi_{i,\vec k}(t)] $ on a de Sitter background
 (the same behaviour is observed in power-law inflation,
 $ a \propto t^p $, $ p > 1 $).
Values of the parameters: $ \gamma = 1/(25\,H^2) $, $ k = 2 H $.
}
\end{figure}

The evolution of the two modes of each degree of freedom can thus be
 easily deduced from \eqref{eq:eoms-vector-dS}:
at the beginning of inflation, when $ \gamma\,H^2\,z^2 \gg 1 $,
 that is when $ \gamma\,(k/a)^2 \gg 1 $ (but with $ \ell_\mathrm{Planck}^2\,(k/a)^2 \ll 1 $ to remain below the transplanckian regime), the two modes oscillate
 with a constant amplitude as in $ M_4 $.
When the ``effective mass'' term $ 1/(\gamma\,H^2\,z^2) $ becomes
 dominant the degree of freedom does not tend to a constant (as all
 degrees of freedom do in standard inflation,
 see Fig.~\ref{fig:standard} in Appendix~\ref{app:standard});
on the contrary both modes oscillate more and more rapidly
 as $ z \to 0 $, albeit with a decreasing amplitude:
\begin{equation}
\bar\Psi_i
\propto
  e^{\pm i z}
\quad\text{for}\quad
\gamma\,H^2\,z^2
\gg
  1\,;
\quad
\bar\Psi_i
\propto
  \sqrt z\,e^{\pm i \ln z/(\sqrt\gamma H)}
\quad\text{for}\quad
\gamma\,H^2\,z^2
\ll
  1\,.
\label{eq:sol-vector}
\end{equation}
See Fig.~\ref{fig:vector}.

Here it may be worthwhile to note that if one considers the case $ \gamma\,H^2 \gg 1 $, the asymptotic behaviour of the vector modes become $ \bar\Psi_i \propto z^{1/2\pm\nu} $, where $ \nu \approx 1/2 - 1/(\gamma\,H^2) $ as $ z \to 0 $.

In the general case now, that is for generic $ a(\eta) $,
 equation \eqref{eq:eoms-vector} reads
\begin{equation}
\bar\Psi_i''
+ \left(k^2 + \frac{a^2}{\gamma}\right)\,\bar\Psi_i
= 0
\end{equation}
whose asymptotic zero-mode solutions are given in the WKB approximation by
\begin{equation}
\bar\Psi_i
\propto
  \frac{1}{\sqrt a}\,e^{\pm i t/\sqrt\gamma}
\end{equation}
hence showing the generality of the result \eqref{eq:sol-vector}.

In the newtonian gauge $ \bar\Psi_i $ identifies to the metric perturbation $ \bar B_i $.
In any other gauge we have $ \bar\Psi_i = \bar B_i - \bar E'_i $, see Eq.~\eqref{eq:gi}, where either $\bar B_i$ or $\bar E_i$ are chosen at will.

\subsection{
Tensor perturbations
}

\begin{figure}
\includegraphics[scale=1.0]{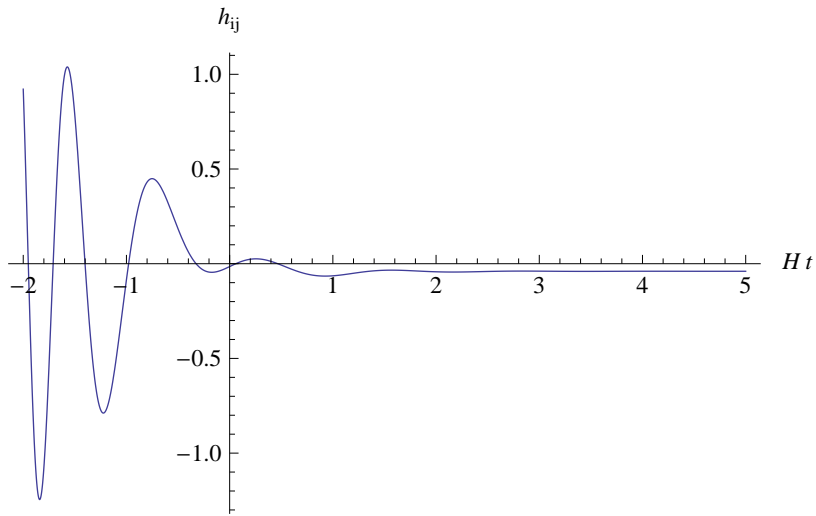}
\caption{\label{fig:tensor}
Evolution in cosmic time of a Fourier component of the tensor metric perturbation $ \bar h_{ij} $ on a de Sitter background (the same behaviour is observed in power-law inflation).
Values of the parameters: $ \gamma = 1/(25\,H^2) $, $ k = 2 H $.
}
\end{figure}

Let us now turn to the Eq.~\eqref{eq:eoms-tensor} for the two tensor perturbations $ \bar h_{ij} $.
It is a fourth order differential equation which hence describes no longer two, as in standard inflation, but four degrees of freedom (just like in flat spacetime, see \cite{Stelle,Deruelle}).
During inflation when spacetime can be described by a de Sitter space with $ a = 1/(-H\,\eta) $, the exact solution of Eq.~\eqref{eq:eoms-tensor} is known \cite{Clunan}.
It can be written, in Fourier space, as
\begin{equation}
\bar h_{ij}
= \frac{1}{a} (\bar\mu_{ij}^{(\mathrm E)} + \bar\mu_{ij}^{(\mathrm W)})
\quad\text{with}\quad
\left\{
\begin{aligned}
&
\frac{d^2}{dz^2} \bar\mu_{ij}^{(\mathrm E)}
+ \left(1 - \frac{2}{z^2}\right)\,\bar\mu_{ij}^{(\mathrm E)}
= 0\,, \\
&
\frac{d^2}{dz^2} \bar\mu_{ij}^{(\mathrm W)}
+ \left(1 + \frac{1}{\gamma\,H^2\,z^2}\right)\,\bar\mu_{ij}^{(\mathrm W)}
= 0\,.
\end{aligned}
\right.
\end{equation}
Hence the two ``Einstein'' degrees of freedom $ \bar h_{ij}^{(\mathrm E)} \propto e^{\pm i z}\,(1 \mp i z) $, which are the same as in standard inflation, first oscillate, decreasing as $ z $, and then tend to constants when $ z \ll 1 $, see Fig.~\ref{fig:standard} of Appendix~\ref{app:standard}.
As for the two Weyl degrees of freedom $ \bar\mu_{ij}^{(\mathrm W)} $, they behave like the two vector degrees of freedom and always oscillate, see Eq.~\eqref{eq:eoms-vector-dS} and Fig.~\ref{fig:vector}.
Hence the amplitude of $ \bar h_{ij}^{(\mathrm W)} = \bar\mu_{ij}^{(\mathrm W)}/a $, first decreases as $ z $, and then as $ z^{3/2} $ as $ z \to 0 $.
Therefore the full Fourier component of the metric perturbation $ \bar h_{ij} $
 first oscillates and, as inflation progresses,
 the standard Einstein modes eventually dominate, tending
 to a constant, see Fig.~\ref{fig:tensor}.

In power-law inflation, $ a(t) \propto t^p $ with $ p > 1 $, where $ t $ is cosmic time,
 the equation of motion \eqref{eq:eoms-tensor} does not split into two second order
 differential equations, one for the Einstein degree of freedom the other for
 the Weyl ghost, as when the background is de Sitter spacetime.
Its solutions however behave similarly, see Fig.~\ref{fig:tensor}.

More generally we find that the two Einstein zero-modes, solutions of $ \bar h_{ij}'' + 2 \mathcal H\,\bar h_{ij}' = 0 $ are approximate solutions of the full zero-mode equation of motion, $ \bar h_{ij}'' + 2 \mathcal H\,\bar h_{ij}' + \gamma\,a^{-2}\,\bar h_{ij}^{(4)} = 0 $, if $ \gamma\,H^2 \ll 1 $ (with $ H = \mathcal H/a $).
As for two Weyl zero-modes they can be found using the WKB approximation so that, all in all, the four independent tensorial zero-modes behave as
\begin{equation}
\bar h_{ij}
\propto
  \left\{
   1\,,
   \quad
   \int^t\!\frac{dt}{a^3}\,,
   \quad
   a^{-3/2}\,e^{i t/\sqrt\gamma}\,,
   \quad
   a^{-3/2}\,e^{-i t/\sqrt\gamma}
  \right\}
\end{equation}
hence confirming that as inflation progresses a generic linear combination of the four modes will tend to a constant.

\subsection{
Scalar perturbations: a master equation and its solutions
}

The analysis of the three Eq.~\eqref{eq:eoms-scalar} for the scalar perturbations is slightly more involved.
However it is easy to extract from them a master equation for $ W $, which reads, in Fourier space:
\begin{equation}
\gamma\,
\left(W^{(4)} - \frac{\ddot H}{\dot H} W^{(3)}\right)
+ C_2\,\ddot W
+ C_1\,\dot W
+ C_0\,W
= 0\,,
\label{eq:eom-master}
\end{equation}
where $ W = \Psi_\mathrm n - \Phi_\mathrm n $, where a dot means derivation with respect to cosmic time $ t = \int^\eta\!a\,d\eta $, where $ H \equiv \dot a/a $, and with
\begin{equation}
\left\{
\begin{aligned}
C_2
&
= 1 + \gamma \left(\frac{2 k^2}{a^2} - H^2\right)\,, \\
C_1
&
= H
  - \frac{\ddot H}{\dot H}
  + \gamma\,
    \left[
     \frac{H^2\ddot H}{\dot H}
     - 3 H\,\dot H
     - \frac{k^2}{a^2}
       \left(\frac{\ddot H}{\dot H} + 4 H\right)
    \right]\,, \\
C_0
&
= 2 \dot H - \frac{H\,\ddot H}{\dot H} + \frac{k^2}{a^2}
  + \frac{\gamma k^2}{a^2}
    \left(
     \frac{H\,\ddot H}{\dot H}
     - \frac{4}{3} \dot H
     + 2 H^2
     + \frac{k^2}{a^2}
    \right)\,.
\end{aligned}
\right.
\end{equation}
(When $ \gamma = 0 $, we have that $ W = -2 \Phi_\mathrm n $ (see Eq.~\eqref{eq:eoms-scalar}) and Eq.~\eqref{eq:eom-master} reduces as it must to the equation for $ \Phi_\mathrm n $ in standard inflation given in Appendix~\ref{app:standard}, Eq.~\eqref{eq:eom-master-standard}.)

Equation \eqref{eq:eom-master} is fourth order and therefore describes two degrees of freedom, and not only one as in standard inflation (see Appendix~\ref{app:standard}).
Ideally one should try to decompose \eqref{eq:eom-master} into two second order differential equations for an ``Einstein'' and a ``Weyl'' degree of freedom (as was done in \cite{Clunan} for the tensorial perturbations on a de Sitter background, see above).
We leave this to further work and content ourselves here with an analysis of the solutions of \eqref{eq:eom-master}.

To have a grip of their behaviour we assume power-law inflation: $ a(t) \propto t^p $ with $ p > 1 $.

Proceeding along the line that we followed to analyse the tensor modes we find that the standard inflationary zero-modes which solve Eq.~\eqref{eq:eom-master} when $ \gamma = 0 $ approximately solve the full equation if $ \gamma\,H^2 \ll 1 $.
As for the other two modes they are found using the WKB approximation, so that the four independent zero modes of Eq.~\eqref{eq:eom-master} have the following late time behaviour
\begin{equation}
W
\propto
  \{
   1\,,
   \quad
   t^{-(1+p)}\,,
   \quad
   t^\frac{p}{2}\,e^{i t/\sqrt\gamma}\,,
   \quad
   t^\frac{p}{2}\,e^{-i t/\sqrt\gamma}
  \}\,.
\label{eq:sol-master}
\end{equation}
(This is confirmed by an analysis of the leading behaviour of the solutions at the irregular singular point at infinity, as well as the exact zero mode solutions which can be written in terms of Bessel and hypergeometric functions.)

These behaviours are in striking contrast to those of the tensor modes which are dominated by the constant, Einstein-mode.
Here both Einstein modes are subdominant.

\begin{figure}
\includegraphics[scale=1.0]{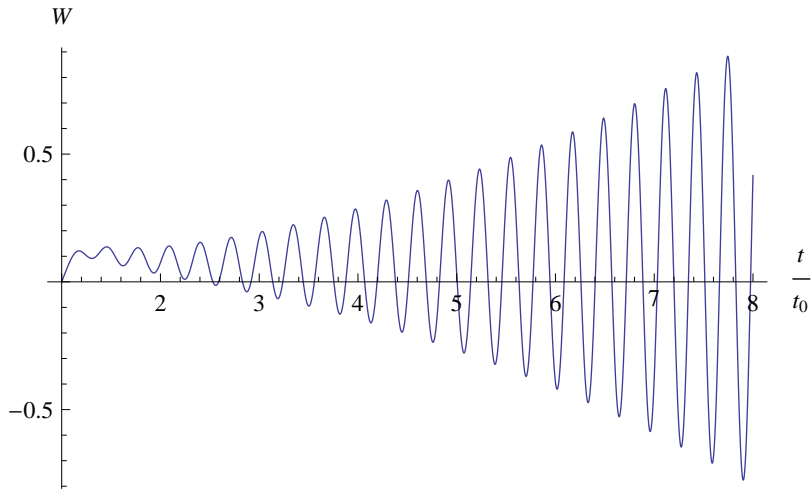}
\caption{\label{fig:scalar}
Evolution in cosmic time of a Fourier component of the scalar mode $ W $
 in power-law inflation.
Values of the parameters: $ a = a_0\,(t/t_0)^p $, $ H_0 = p/t_0 $ with $ p = 4 $, and $ \gamma = 1/(25\,H_0^2) $, $ k = 2 H_0\,a_0 $;
initial conditions at $ t = t_0 $: $ W = 0 $, $ \dot W = 1/t_0 $, $ \ddot W = 0 $, $ W^{(3)} = 1/t_0^3 $.
}
\end{figure}

The evolution of a typical Fourier component of $ W $ is given in Fig.~\ref{fig:scalar}.
As one can see, not only does the Fourier component $ W $ never ``freeze out'' but its amplitude \emph{increases} as inflation proceeds instead of tending to a constant as in standard inflation when $ \gamma = 0 $ (see Appendix~\ref{app:standard}).

Equation \eqref{eq:eom-master} and the behaviour \eqref{eq:sol-master} of its modes can be seen as the main result of this paper since, knowing $ W $, we can predict the behaviour of all cosmological perturbations of the model.

\subsection{
Evolution of the scalar perturbations in the newtonian gauge
}

Once $ W $ is known, $ \Phi_\mathrm n $ and $ \chi_\mathrm n $ follow from Eq.~\eqref{eq:eoms-scalar}:
\begin{equation}
\left\{
\begin{aligned}
-2 \Phi_\mathrm n
&
= W + \gamma\,\left(\ddot W + H\,\dot W + \frac{k^2}{3 a^2} W\right), \\
\kappa\,\dot\varphi\,\chi_\mathrm n
&
= \dot W
  + H\,W
  + \gamma\,
    \left[
     W^{(3)} + \dot W\,(\dot H - H^2) + \frac{k^2}{a^2} (\dot W - H\,W)
    \right]\,.
\end{aligned}
\right.
\label{eq:Psichi}
\end{equation}

It follows from \eqref{eq:sol-master} and \eqref{eq:Psichi} that $ \Psi_\mathrm n $ increases as $ W $ and that $ \Phi_\mathrm n $ and $ \chi_\mathrm n $ behave as
 $ t^{p/2 - 1}\,e^{\pm i t/\sqrt\gamma} $.
(The leading term in $ \Phi_\mathrm n $ grows a priori like $ W $ but cancels out; as for $ \dot\varphi\,\chi_\mathrm n \propto \chi_\mathrm n/t $ it should also a priori grow like $ W $ but the two first leading orders cancel out.)

Since $ \Phi_\mathrm n = C $, $ \Psi_\mathrm n = A $ and $ \chi_\mathrm n = \delta\phi $ in the newtonian gauge, we therefore reach the conclusion that in that gauge all cosmological perturbations blow up in single field inflation with a Weyl term.

This result does not mean however that they blow up in all coordinate systems, as we see now.

\subsection{
Evolution of the scalar perturbations in the comoving slicing
}

All linear combinations of $ \Phi_\mathrm n $, $ \Psi_\mathrm n $ and $ \chi_\mathrm n $ are gauge invariant and can be expressed in terms of $ W $ only.
Moreover, using Eq.~\eqref{eq:gi}, they give the perturbations ($ A $, $ B $, $ C $, $ E $) of the metric and the perturbation $ \delta\phi $ of the scalar field in any gauge.
Therefore gauge invariant quantities can be built which identify to various perturbations in a given gauge.

An example a such a gauge invariant perturbation is the curvature perturbation \cite{Kodama}:
\begin{equation}
\mathcal R_\mathrm c
\equiv
  \Phi_\mathrm n - \frac{\mathcal H\,\chi_\mathrm n}{\varphi'}
= C - \frac{\mathcal H\,\delta\phi}{\varphi'}
\label{eq:def-R}
\end{equation}
(introduced in standard inflation in various guise, see \cite{Bardeen2,Kodama,Mukhanov} and Appendix~\ref{app:standard}).
In the ``comoving slicing'' gauge, that is, in the coordinate system where
\begin{equation}
\delta\phi
= 0,
\end{equation}
$ \mathcal R_\mathrm c $ (denoted by $ -\zeta $ in \cite{Mukhanov})
 identifies with the metric perturbation $ C $.

From its definition \eqref{eq:def-R} one expects a priori that $ \mathcal R_\mathrm c $ should grow like $ \Phi_\mathrm n $, that is as $ t^{p/2-1}\,e^{\pm i t/{\sqrt\gamma}} $ or perhaps at a slower rate if the first leading terms cancel out, but it happens that the cancellation is so drastic as to make $ \mathcal R_\mathrm c $ not to grow (despite the fact that $ p $ can be as big as one wishes).

Indeed, $ \mathcal R_\mathrm c $, using \eqref{eq:Psichi}, is a function of $ W $ up to $ W^{(3)} $.
If we compute $ \dot{\mathcal R}_\mathrm c $ and use the master equation \eqref{eq:eom-master} to eliminate $ W^{(4)} $ we find that
\begin{equation}
\dot{\mathcal R}_\mathrm c
= \frac{k^2}{2 a^2}
  \left\{
   - \frac{H}{\dot H} W
   + \gamma\,
     \left[
      \left(\frac{2}{3} + \frac{H^2}{\dot H}\right) \dot W
      - \frac{H}{\dot H} \left(\ddot W + \frac{k^2}{a^2} W\right)
     \right]
  \right\}\,.
\end{equation}
(When $ \gamma = 0 $ we recover the well-known result of standard inflation, see Appendix~\ref{app:standard}.)

The analytic analysis of $ \dot{\mathcal R}_\mathrm c $, as well as numerical
 plots show without any ambiguity that it goes to zero as
 $ t^{-3p/2}\,e^{\pm i t/\sqrt\gamma} $.
We must therefore conclude that indeed all leading terms up to order
 $ t^{-3p/2}\,e^{\pm i t/\sqrt\gamma} $ do cancel out when looking carefully at 
 the asymptotic behaviour of $ \mathcal R_\mathrm c $ from its definition.
More precisely, knowing the asymptotic behaviour of the four independent solutions
 for $ W $, see \eqref{eq:sol-master} we have that
\begin{equation*}
\dot{\mathcal R}_\mathrm c
\propto
  \{
   t^{1-2p}\,,
   \quad
   t^{-3p}\,,
   \quad
   t^{-3 p/2}\,e^{i t/\sqrt\gamma}\,,
   \quad
   t^{-3 p/2}\,e^{-i t/\sqrt\gamma}
  \}\,.
\end{equation*}

Consider now the following gauge invariant perturbation:
\begin{equation}
A_\mathrm c
\equiv
  \Psi_\mathrm n - \left(\frac{\chi_\mathrm n}{\dot\varphi}\right)^.
= A - \left(\frac{\delta\phi}{\dot\varphi}\right)^.
\end{equation}
which identifies to the metric perturbation $ A $ in comoving slicing.

Again one expects a priori from its definition that $ A_\mathrm c $ will grow, like $ \Psi_\mathrm n $ that is like $ W $ if the leading orders do not cancel out.
But, again, this not the case.
Indeed $ A_\mathrm c $ depends, using \eqref{eq:Psichi}, on $ W $ and its derivatives up to the fourth.
Using the master equation \eqref{eq:eom-master} to eliminate $ W^{(4)} $ we have that
\begin{equation}
A_\mathrm c
= -\frac{k^2}{2 a^2\,\dot H}\,
  \left[
   \gamma\,(\ddot W - H\,\dot W)
   + W\,\left(1 + \frac{\gamma\,k^2}{a^2}\right)
  \right]\,.
\end{equation}
We also have that
\begin{equation}
\dot{\mathcal R}_\mathrm c - H\,A_\mathrm c
= \frac{\gamma k^2}{3 a^2} \dot W\,.
\end{equation}

Knowing the asymptotic behaviour of $ W $, see \eqref{eq:sol-master}, we see that the growing modes solve the equation $ \gamma\,(\ddot W - H\,\dot W) + W = 0 $.
Therefore the $ W $ mode which gives the asymptotic behaviour of $ A_\mathrm c $ is the subdominant mode $ W = 1 $.
Hence $ A_\mathrm c $ decays to zero as $ t^{2-2p} $, without oscillating (contrarily to $ \mathcal R_\mathrm c $), as numerical plots confirm.

We therefore have shown that the metric perturbations $ A $ and $ C $, do \emph{not} grow when evaluating them in the comoving slicing.
This is in striking contrast to their behaviour in the newtonian gauge, where they blow up, as we have seen in the previous section.

A last check has to be done though, since in the comoving slicing the metric perturbation $ (B-E') $ is not zero.
To find it one uses the expression for $ \chi_\mathrm n $ in terms of $ \delta\phi $: $ \chi_\mathrm n = \delta\phi + \varphi'\,(B-E') $, see Eq.~\eqref{eq:gi2}.
In the comoving slicing $ \delta\phi = 0 $.
Therefore $ B - E' = \chi_\mathrm n/\varphi' = \chi_\mathrm n/(a\,\dot\varphi) $.
We know from the previous section the asymptotic behaviour of $ \chi_\mathrm n $: $ \chi_\mathrm n \propto t^{p/2-1}\,e^{\pm i t/{\sqrt\gamma}} $.
Therefore $ (B-E') $ also decays, as $ t^{-p/2}\,e^{\pm i t/{\sqrt\gamma}} $.

\section{
Conclusions
}

In this paper, we have studied the role of the Weyl term on the evolution of linear cosmological perturbations when the Friedmann-Lema\^itre background is that of single field inflation.

We found that the two Weyl tensor degrees of freedom are tamed by the Einstein gravitational waves and thus do not spoil too much the evolution of the tensorial cosmological perturbations as given by the standard inflationary scenario, whether the background is approximated by a de Sitter spacetime as in \cite{Clunan} or in power-law inflation, see Fig.~\ref{fig:tensor}.
Vector modes on the other hand, which are absent in standard inflation, do propagate; these two pure-Weyl vector degrees of freedom never ``freeze out'', but their amplitude decreases as inflation proceeds, see Fig.~\ref{fig:vector}.
Finally, the evolution of the scalar modes is drastically modified by the presence of the Weyl term: instead of one there are now two scalar degrees of freedom which, when working in the newtonian gauge, and contrarily to what happens in standard inflation, not only do not freeze out but their amplitude increases during inflation, see Fig.~\ref{fig:scalar}.
However there exists at least one coordinate system (the comoving slicing) where none of the perturbations grows.

We cannot therefore claim at this stage whether the five Weyl degrees of freedom, which are ghosts in Minkowski spacetime, screw up or not the evolution of linear cosmological perturbations in inflation, since their asymptotic behaviour depends crucially on the coordinate system used.

To complete our study, and arrive at a more definite conclusion, an hamiltonian analysis of the action (\ref{eq:action-pert}--\ref{eq:action-pert-Einstein}) should be performed to isolate the Weyl degrees of freedom from the Einstein's ones.
It is however clear from the form of the action that the vector perturbations $ \bar\Psi_i $ are two ghosts, since the sign of their kinetic term is positive; it is also clear that their quantisation will impose a normalisation of their Fourier modes in $ k^{-3/2} $ because of the presence of the extra spatial derivatives in the kinetic term in the action $ S^{(\mathrm V)} $.
As for the tensor perturbations they were analysed in \cite{Clunan} when the background is approximated by a de Sitter spacetime: the two Weyl degrees of freedom $ \bar\mu_{ij}^{(\mathrm W)} $ are ghosts, and the normalisation of the Einstein modes is modified by their presence.
It remains however to generalize this analysis to the case when the background is no longer de Sitter spacetime.
Finally, the hamiltonian analysis of the action $ S^{(\mathrm S)} $ (\ref{eq:action-pert}--\ref{eq:action-pert-Einstein}) for the scalar perturbations, in order to isolate the ghost degree of freedom, is more tricky and is left to further work, see \cite{Deruelle}.

To complete our study the question of what happens at the end of inflation should also be addressed.
As a first step this transitory period could be modelled by a sudden transition from the inflationary stage with $ a \propto t^p $ with $ p > 1 $ to the radiation era $ a \propto t^{1/2} $.
The junction conditions which give the perturbations after the transition in terms of their behaviour during inflation are well-known in Einstein's theory, see \cite{Deruelle2}.
When the Weyl-term is present they have to be analysed anew since the equations of motion become fourth order.
One expects however that the two tensorial Weyl ghosts will not change too much the standard picture since they are subdominant compared to Einstein's gravitational waves.
The matching of the two ghost vector degrees of freedom, although decaying during inflation, may on the other hand be more tricky as they oscillate at very high frequency at the end of inflation.
Finally a proper matching of the two scalar degrees of freedom requires first the hamiltonian analysis referred to above.

Last but not least a complete study of the role of Weyl's ghosts in inflation requires an analysis of observables, such as the CMB temperature fluctuations, which may be affected by their presence.

In any case we have already seen in this paper that the addition of the Weyl term to the action of Einstein's gravity coupled to a scalar field modifies drastically the evolution of perturbations in inflationary cosmological models and we gave some of the necessary tools to assess their influence in observational cosmology.

\begin{acknowledgments}
N.D.\ and Y.S.\ thank the Yukawa Institute for its hospitality when this work was completed.
N.D.\ also acknowledges financial support from the CNRS-JSPS contract 24600.
Y.S.\ was supported in part by JSPS Postdoctoral Fellowships for Research Abroad.
M.S.\ is supported by Korea Institute for Advanced Study under the KIAS Scholar program.
This work was supported in part by JSPS Grant-in-Aid for Scientific
Research (A) No.~18204024, MEXT Grant-in-Aid for Creative Scientific Research
No.~19GS0219, and MEXT Grant-in-Aid for the Global COE programs
``The Next Generation of Physics, Spun from Universality and Emergence''
at Kyoto University.
\end{acknowledgments}
\newpage
\appendix

\section{\label{app:pert}
Perturbed quantities
}

\subsection{
Perturbed Weyl and Bach tensors (See definitions in footnote~\ref{fn:def})
}

\begin{itemize}
\item
\textsl{Linearised Weyl tensor on $ M_4 $}:

\begin{itemize}
\item
Scalar part (with $ W \equiv \Psi_\mathrm n - \Phi_\mathrm n $):
\begin{equation}
\left\{
\begin{aligned}
C^{(\mathrm S)}_{0i0j}
&
= \frac{1}{2} \partial_{ij} W
  - \frac{1}{6} \delta_{ij}\,\triangle W\,, \\
C^{(\mathrm S)}_{0ijk}
&
= 0\,, \\
C^{(\mathrm S)}_{ijkl}
&
= \frac{1}{2}
  (\delta_{ik}\,\partial_{jl}
   - \delta_{il}\,\partial_{jk}
   - \delta_{jk}\,\partial_{il}
   + \delta_{jl}\,\partial_{ik})\,
  W
  - \frac{1}{3}
    (\delta_{ik}\,\delta_{jl}
     - \delta_{il}\,\delta_{jk})\,
    \triangle W\,.
\end{aligned}
\right.
\end{equation}

\item
Vector part:
\begin{equation}
\left\{
\begin{aligned}
C^{(\mathrm V)}_{0i0j}
&
= \frac{1}{4}
  (\partial_i \dot{\bar\Psi}_j + \partial_j \dot{\bar\Psi}_i)\,, \\
C^{(\mathrm V)}_{0ijk}
&
= \frac{1}{2}
  \partial_i (\partial_j \bar\Psi_k - \partial_k \bar\Psi_j)
  - \frac{1}{4}
    (\delta_{ij}\,\triangle \bar\Psi_k - \delta_{ik}\,\triangle \bar\Psi_j)\,, \\
C^{(\mathrm V)}_{ijkl}
&
= \frac{1}{4}
  [\delta_{ik}\,(\partial_j \dot{\bar\Psi}_l + \partial_l \dot{\bar\Psi}_j)
   - \delta_{il}\,(\partial_j \dot{\bar\Psi}_k + \partial_k \dot{\bar\Psi}_j)
   - \delta_{jk}\,(\partial_i \dot{\bar\Psi}_l + \partial_l \dot{\bar\Psi}_i)
   + \delta_{jl}\,(\partial_i \dot{\bar\Psi}_k + \partial_k \dot{\bar\Psi}_i)]\,.
\end{aligned}
\right.
\end{equation}

\item
Tensor part:
\begin{equation}
\left\{
\begin{aligned}
C^{(\mathrm T)}_{0i0j}
&
= -\frac{1}{4} \ddot{\bar h}_{ij} - \frac{1}{4} \triangle \bar h_{ij}\,, \\
C^{(\mathrm T)}_{0ijk}
&
= -\frac{1}{2} \partial_j \dot{\bar h}_{ik}
  + \frac{1}{2} \partial_k \dot{\bar h}_{ij}\,, \\
C^{(\mathrm T)}_{ijkl}
&
= \frac{1}{2}
  (-\partial_{ik} \bar h_{jl}
   + \partial_{il} \bar h_{jk}
   + \partial_{jk} \bar h_{il}
   - \partial_{jl} \bar h_{ik})
  + \frac{1}{4}
    (\delta_{ik}\,\square \bar h_{jl}
     - \delta_{il}\,\square \bar h_{jk}
     - \delta_{jk}\,\square \bar h_{il}
     + \delta_{jl}\,\square \bar h_{ik})\,.
\end{aligned}
\right.
\end{equation}
\end{itemize}

Since the perturbed Friedmann-Lema\^itre metric is conformal to the perturbed Minkowski metric, and as can be checked explicitly, the
components of $ C^\mu{}_{\nu\rho\sigma} $ are the same for both metrics.

\item
\textsl{Linearised Bach tensor on $ M_4 $}:

At linear order around $ M_4 $ the Bach tensor reduces to $ B_{\mu\nu} = \partial^{\rho\sigma} C_{\mu\rho\nu\sigma} $.
Its components are:
\begin{equation}
\left\{
\begin{aligned}
B^{(\mathrm S)}_{00}
&
= \frac{1}{3} \triangle \triangle W\,,
\quad
B^{(\mathrm S)}_{0i}
= \frac{1}{3} \partial_i \triangle W'\,,
\quad
B^{(\mathrm S)}_{ij}
= \frac{1}{2} \partial_{ij}
  \left(W'' - \frac{1}{3} \triangle W\right)
  - \frac{1}{6} \delta_{ij}\,(W'' - \triangle W)\,, \\
B^{(\mathrm V)}_{00}
&
= 0\,,
\quad
B^{(\mathrm V)}_{0i}
= -\frac{1}{4} \triangle \square \bar\Psi_i\,,
\quad
B^{(\mathrm V)}_{ij}
= -\frac{1}{4}
  (\partial_i \square \bar\Psi_j' + \partial_j \square \bar\Psi_i')\,, \\
B^{(\mathrm T)}_{00}
&
= 0\,,
\quad
B^{(\mathrm T)}_{0i}
= 0\,,
\quad
B^{(\mathrm T)}_{ij}
= -\frac{1}{4} \square \square \bar h_{ij}\,.
\end{aligned}
\right.
\end{equation}

The linearised components of $ B_{\mu\nu} $ on a Friedmann-Lema\^itre metric are the same, up to a factor $ 1/a^2 $.
\end{itemize}

\subsection{
Weyl action at quadratic order
}

Using the expressions for the linearised Weyl tensor on $ M_4 $ given above and because of the conformal invariance of the Weyl action we have
\begin{equation}
\begin{aligned}
&
\frac{1}{4} \int\!d^4x\,
(\sqrt{-g}\,
C_{\mu\nu\rho\sigma}\,C^{\mu\nu\rho\sigma})_{\text{perturbed FL}}
= \frac{1}{4} \int\!d^4x\,
  (\sqrt{-g}\,
   C_{\mu\nu\rho\sigma}\,C^{\mu\nu\rho\sigma})_{\text{perturbed Minkowski}} \\
&
= \int\!d^4x\,
  \left[
   \frac{1}{3} (\triangle W)^2
   + \frac{1}{4}
     (\partial_i \dot{\bar\Psi}_j\,\partial^i \dot{\bar\Psi}^j
      - \triangle \bar\Psi_i\,\triangle \bar\Psi^i)
   + \frac{1}{8}
     (\ddot{\bar h}_{ij}\,\ddot{\bar h}^{ij}
      - 2 \partial_k \dot{\bar h}_{ij}\,\partial^k \dot{\bar h}^{ij}
      + \triangle \bar h_{ij}\,\triangle \bar h^{ij})
  \right]\,.
\end{aligned}
\end{equation}

\section{\label{app:fromaction}
The equations of motion from the action for gauge invariant perturbations
}

NB: in this appendix we suppress the index $ \mathrm n $ which ornate $ \Phi $ and $ \Psi $ and $ \chi $ in the main text.

Consider the action $ S^{(\mathrm S)} $ in terms of the three perturbations $ \Phi $, $ \Psi $ and $ \chi $ given in \eqref{eq:action-pert} \eqref{eq:action-pert-Einstein}.

Extremisation of $ S^{(\mathrm S)} $ with respect to $ \Psi $ gives the $ (00) $-component of the perturbed equations of motion, already obtained in \eqref{eq:eoms-scalar}, that is:
\begin{equation}
6 \mathcal H\,\Phi'
- 2 \triangle \Phi
- (\varphi'\chi' + 2 a^2\,V\,\Psi + a^2\,V_{,\varphi}\,\chi)
= \frac{2 \gamma}{3 a^2} \triangle \triangle (\Psi - \Phi)\,.
\label{eq:eom-scalar1}
\end{equation}
Extremisation of $ S^{(\mathrm S)} $ with respect to $ \chi $ gives the perturbed Klein-Gordon equation
\begin{equation}
\chi''
+ 2 \mathcal H\,\chi'
- \triangle \chi
+ a^2\,V_{,\varphi\varphi}\,\chi
- \varphi'\,(\Psi' - 3 \Phi')
+ 2 a^2\,V_{,\varphi}\,\Psi
= 0
\label{eq:eom-scalar2}
\end{equation}
Finally, extremisation of $ S^{(\mathrm S)} $ with respect to $ \Phi $ gives
\begin{equation}
\triangle
\left(\Psi + \Phi - \frac{\gamma}{3 a^2} \triangle (\Psi - \Phi)\right)
= \frac{3}{a^2} (a^2 F)'
\quad\text{with}\quad
F
\equiv
  \frac{\kappa\,\varphi'\,\chi}{2} + (\Phi' - \mathcal H\,\Psi)\,.
\label{eq:eom-scalar3}
\end{equation}
Let us now perform the following manipulations:
\begin{enumerate}
\item
Replace in \eqref{eq:eom-scalar1} $ \chi $ and $ \chi' $ by their expressions in function of $ F $ given in \eqref{eq:eom-scalar3}; \eqref{eq:eom-scalar1} then depends on $ \Psi $, $ \Phi $ and $ F $.

\item
Replace $ \Psi $ by its expression in terms of $ F $ and $ \Phi $ using \eqref{eq:eom-scalar3};
\eqref{eq:eom-scalar1} then depends on $ \Phi $ and $ F $ only.

\item
Extract from it the expression of $ \Phi'' $ and compute its time derivative $ \Phi''' $.
\end{enumerate}

Consider now Eq.~\eqref{eq:eom-scalar2}: after the above replacements of $ \chi $ and $ \Psi $ it depends on $ F $, $ \Phi $ and its time derivatives up to the third.
Replace $ \Phi'' $ and $ \Phi''' $ by the expressions obtained previously and find (we used Mathematica) that it can be written as
\begin{equation}
\Delta G
= 0
\quad\text{with}\quad
G
\equiv
  \frac{\kappa\,\varphi'\,\chi}{2}
  + (\Phi' - \mathcal H\,\Psi)
  + \frac{\gamma}{3 a^2} \triangle (\Psi - \Phi)'\,.
\end{equation}
Therefore the Klein-Gordon equation \eqref{eq:eom-scalar2} is equivalent to the $ (0i) $-scalar component of the equations of motion \eqref{eq:eoms-scalar}.

As for \eqref{eq:eom-scalar3}, since $ G = 0 $ and hence $ F = -\frac{\gamma}{3 a^2} \triangle (\Psi - \Phi)' $ with $ F $ given in \eqref{eq:eom-scalar3}, it becomes the $ (ij) $-scalar components ($ i \neq j $) of the equations of motion \eqref{eq:eoms-scalar}.

This (simple) derivation must be contrasted to those found in the literature.
In \cite{Mukhanov} and \cite{Makino} for example, $ \kappa\,S^{(\mathrm S)}_\mathrm E $ is computed without specifying the coordinate system, that is for the full metric
\begin{equation*}
ds^2
= a(\eta)^2\,
  \{
   -(1+2A)\,d\eta^2
   + 2 \partial_i B\,dx^i\,d\eta
   + [\delta_{ij}\,(1 + 2 C) + 2 \partial_{ij} E]\,dx^i\,dx^j
  \}
\end{equation*}
and the action (in the case $ \gamma = 0 $) hence gets an extra-term, proportional to $ (B - E')\,F $, with $ F \equiv \frac{\kappa\,\varphi'\,\chi}{2} + (\Phi' - \mathcal H\,\Psi) $.

Extremisation with respect to $ (B - E') $ gives $ F = 0 $, that is, the $ (0i) $-component of the equations of motion given in \eqref{eq:eoms-scalar} (for $ \gamma = 0 $).
Extremisation with respect to $ \Phi $, Eq.~\eqref{eq:eom-scalar3}, then gives the last equations of motion \eqref{eq:eoms-scalar}.
(And the perturbed Klein-Gordon equation is ignored.)

\section{\label{app:standard}
The evolution of cosmological perturbations during standard inflation: recap
}

In the case of standard inflation ($ \gamma = 0 $),
 Eq.~\eqref{eq:eoms-vector} is a constraint and vector modes are absent:
\begin{equation}
\bar\Psi_i
= 0\,.
\end{equation}
Equation \eqref{eq:eoms-tensor} describes the two degrees of
 freedom of gravitational waves that freely propagate on the
 Friedmann-Lema\^itre background, and can be rewritten as \cite{Grishchuk}:
\begin{equation}
\bar\mu_{ij}'' - \frac{a''}{a} \bar\mu_{ij} - \triangle \bar\mu_{ij}
= 0
\quad\text{with}\quad
\bar\mu_{ij}
\equiv
  a\,\bar h_{ij}\,.
\end{equation}
As for Eq.~\eqref{eq:eoms-scalar}, it consists in two constraints:
\begin{equation}
\Psi_\mathrm n
= -\Phi_\mathrm n
\quad\text{and}\quad
\frac{\kappa\,\varphi'\,\chi_\mathrm n}{2}
= \mathcal H\,\Psi_\mathrm n - \Phi_\mathrm n'
\label{eq:cons}
\end{equation}
which, inserted into the first of Eq.~\eqref{eq:eoms-scalar}, yield a master equation which can be written into various equivalent forms \cite{Bardeen,Mukhanov,Sasaki,Kodama}, e.g.:
\begin{equation}
\Phi_\mathrm n''
+ 2 \left(\mathcal H - \frac{\varphi''}{\varphi'}\right)\,\Phi_\mathrm n'
+ 2 \left(\mathcal H' - \mathcal H\,\frac{\varphi''}{\varphi'}\right)\,
  \Phi_\mathrm n
- \triangle \Phi_\mathrm n
= 0\,,
\label{eq:eom-master-standard}
\end{equation}
that is,
\begin{equation}
u'' - \frac{\theta''}{\theta} u - \triangle u
= 0
\quad\text{with}\quad
u
\equiv
  \frac{a}{\varphi'} \Phi_\mathrm n
\quad\text{and}\quad
\theta
\equiv
  \frac{\mathcal H}{a\,\varphi'}\,.
\label{eq:eom-master-standard2}
\end{equation}
Using the constraint \eqref{eq:cons} this master equation also reads
\begin{equation}
v'' - \frac{z''}{z} v - \triangle v
= 0
\quad\text{with}\quad
z
\equiv
  \frac{a\,\varphi'}{\mathcal H}
\quad\text{and}\quad
v
\equiv
  a\,\left(\frac{\varphi'}{\mathcal H} \Phi_\mathrm n - \chi_\mathrm n\right)
= \frac{2 a}{\kappa\,\varphi'}
  \left[
   \Phi_\mathrm n'
   + \mathcal H\,\left(2 - \frac{\mathcal H'}{\mathcal H^2}\right)\,
     \Phi_\mathrm n
  \right]\,.
\end{equation}

\begin{figure}
\includegraphics[scale=1.0]{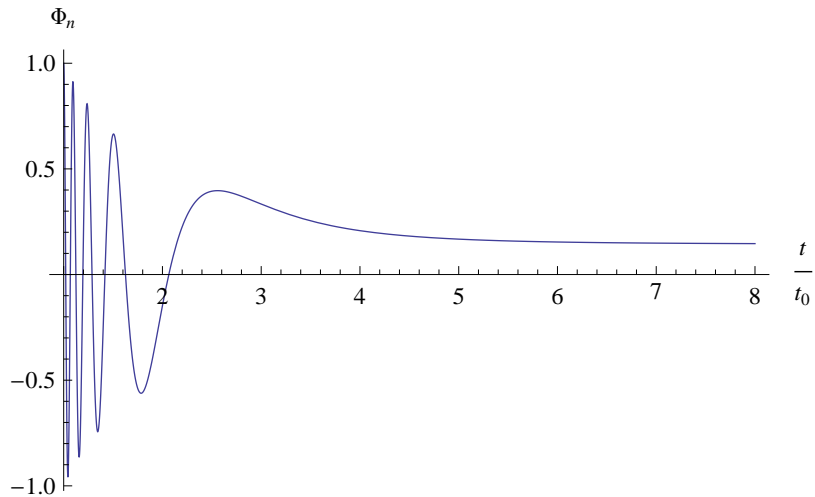}
\caption{\label{fig:standard}
Typical evolution of a cosmological perturbation mode $ \Phi_\mathrm n $ during inflation in Einstein's theory.
}
\end{figure}

The evolution of the two degrees of freedom $ \bar\mu_{ij} $ and the evolution of $ u $ (or $ v $) are similar in the inflationary stage when the ``mass-terms'' $ a''/a $, $ \theta''/\theta $ or $ z''/z $ can be neglected: in Fourier space the two independent modes of each degree of freedom oscillate as in flat spacetime.
When inflation progresses and the mass-terms come to dominate only the dominant modes of $ \bar h_{ij} $ and $ \Phi_\mathrm n $ ``survive'' and become almost constant, see Fig.~\ref{fig:standard}.
More precisely the two zero-modes which solve \eqref{eq:eom-master-standard2} behave as $ \theta $ and $ \theta\,\int^\eta\!d\eta/\theta $.
In the case of power-law inflation, $ a \propto t^p $ this translates as
\begin{equation}
\Phi_\mathrm n
\propto
  \{1\,,\quad t^{-(1+p)}\}\,.
\end{equation}
After the end of inflation when the scale factor increases as $ t^{2/3} $ say, the mass-terms become subdominant again and the modes again oscillate.

Of course other gauge invariant variables can be introduced which, thanks to the constraints \eqref{eq:cons}, can all be expressed in terms of $ \Phi_\mathrm n $.
An example is the curvature perturbation \cite{Kodama}:
\begin{equation}
\mathcal R_\mathrm c
\equiv
  \Phi_\mathrm n - \frac{\mathcal H\,\chi_\mathrm n}{\varphi'}
= \frac{v}{z}
= \Phi_\mathrm n\,
  \left(1 + \frac{2 \mathcal H^2}{\kappa\,\varphi'^2}\right)
  + \frac{2 \mathcal H}{\kappa\,\varphi'^2} \Phi_\mathrm n'
\end{equation}
(denoted by $ -\zeta $ in \cite{Mukhanov}).
It is a useful quantity because its time derivative is given by (using the master equation \eqref{eq:eom-master-standard} to eliminate $ \Phi_\mathrm n'' $):
\begin{equation}
\mathcal R_\mathrm c'
= \frac{2 \mathcal H}{\kappa\,\varphi'^2} \triangle \Phi_\mathrm n\,.
\end{equation}
Therefore the amplification of the $ \Phi_\mathrm n $ modes can easily be obtained from the fact that $ \mathcal R_\mathrm c $ is almost constant as long as the mass-term dominates.

Another interesting gauge invariant perturbation is
\begin{equation}
A_\mathrm c
\equiv
  \Psi_\mathrm n
  - \frac{1}{a} \left(\frac{a\,\chi_\mathrm n}{\varphi'}\right)'
= -\Phi_\mathrm n
  - \frac{1}{a}
    \left[
     \frac{2 a}{\kappa\,\varphi'^2}
     (\mathcal H\,\Phi_\mathrm n + \Phi_\mathrm n')
    \right]'\,,
\end{equation}
which, using \eqref{eq:eom-master-standard}, can be shown to be simply related to $ \mathcal R_\mathrm c $ by: $ \mathcal H\,A_\mathrm c = \mathcal R_\mathrm c $.

All gauge invariant variables can now be related to the perturbations of the metric and the scalar field using $ \Phi_\mathrm n = C + \mathcal H\,(B - E') $ (Eq.~\eqref{eq:gi} in the main text).
Thus, in the newtonian gauge, where $ B = E = 0 $ the gauge invariant perturbations $ \Phi_\mathrm n $, $ \Psi_\mathrm n $ and $ \chi_\mathrm n $ identify respectively to $ C $, $ A $ and $ \delta\phi $.

As for $ \mathcal R_\mathrm c $ and $ A_\mathrm c $ they are related to the metric and scalar field perturbations as
\begin{equation}
\mathcal R_\mathrm c
= C - \frac{\mathcal H\,\delta\phi}{\varphi'}\,,
\quad
A_\mathrm c
= A - \frac{1}{a} \left(\frac{a\,\delta\phi}{\varphi'}\right)'\,.
\end{equation}
In the ``comoving slicing'', that is, in the coordinate system where $ \delta\phi = 0 $, $ \mathcal R_\mathrm c $ and $ A_\mathrm c $ identify to the metric perturbations $ A $ and $ C $.

\bibliographystyle{apsrev}

\end{document}